\documentclass[%
 reprint,
 amsmath,amssymb,
 aps,
]{revtex4-2}

\usepackage{graphicx}
\usepackage{dcolumn}
\usepackage{bm}
\usepackage{cancel}
\usepackage{caption}
\usepackage{subcaption}

\begin{document}

\preprint{APS/123-QED}

\title{Studying neutrino oscillations at DUNE through dynamical Lorentz symmetry breaking in four-Majorana fermion model}

\author{Susie Kim}
\affiliation{%
 Mathematical Institute, University of Oxford, Oxford, UK
}%

\date{\today}

\begin{abstract}
We study the impact of the dynamical Lorentz symmetry breaking induced by the auxiliary gauge fields of neutrino on the oscillations probability at DUNE. The DLSB introduces an alternative energy-momentum relation of the neutrinos and thus results in a new oscillation probability. We extend the previously proposed four-Majorana fermion model that gives rise to DLSB after the type II seesaw mechanism by considering the electron neutrino forward scattering when passing through a medium. Moreover, we incorporate the three-flavor neutrino states, which introduce the CP-violation term inside the oscillation probability. The impact of DLSB parameters around the order of $10^{-2}-10^{-3}$, which are at a strong coupling regime, on the oscillation probability is found to be measurable at DUNE within the 20 years through $\nu_e$ and $\overline{\nu}_e$ disappearance signals. We also compare the predicted spectra of the DLSB oscillations and the oscillation with the CP-violating term equal to $\pi/2$ to conclude that the presence of DLSB would increase the systematic uncertainty for the measurement of CP-violation at DUNE. 
\end{abstract}

\maketitle


\section{\label{sec:level1} Introduction}
One of the great unsolved mysteries of our universe is that of matter-antimatter asymmetry; the majority of our observable universe comprises matter with only a tiny component of antimatter. This phenomenon seems unnatural in the light of the many symmetries which underpin our contemporary understanding of physics. To explain how the early universe gave rise to this imbalance, numerous theories have been proposed. One of the compelling attempts to solve this mystery is to introduce hypothetical right-handed neutrinos (RHN) through the seesaw mechanism. In nature, we only observe left-handed neutrinos (LHN), which have three mass eigenstates with extremely small magnitudes compared to other fermions. The seesaw mechanism is originally proposed to explain the smallness of the LHNs through the existence of RHNs with very large mass~\cite{MINKOWSKI1977421,PhysRevD.20.2986,https://doi.org/10.48550/arxiv.1306.4669}. In one of the plausible scenarios called leptogenesis, the decay of heavy RHNs in the early universe generates the baryon asymmetry. Therefore, the goal of many future neutrino experiments is to verify the principles of the seesaw mechanism and eventually find evidence of leptogenesis. 

There are various theorized models which utilize the seesaw mechanism as an explanation for neutrino mass.  Y. M. P. Gomes and M. J. Neves propose a novel fermion model with two Majorana spinors which we will use as the basis for our analysis~\cite{Gomes}. The two Majorana fermions gain masses through the type II seesaw mechanism, in which the left-handed fermion is assigned a light mass and the right-landed fermion is assigned a heavy mass. The self-interaction of the two Majorana particles of this model introduces 4-vectors as the vacuum expectation value (VEV), which indicates dynamical Lorentz symmetry breaking (DLSB). The appearance of DLSB modifies the energy-momentum relation of the neutrinos and, thus, the neutrino oscillation probability. The transition probability of electron neutrino to muon neutrino in the presence of DLSB at short baseline (SBL) neutrino experiments are evaluated in the literature with the stated goal of reconciling the anomalous oscillation data from LSND with that of Super-Kamiokande. After plotting the parameter space of this model, Gomes and Neves suggest that a DLSB value of $a_{21} \sim \pm 6 \sqrt{3} \pi \times 10^{-19}$ can effectively explain results in both LSND and Super-Kamiokande data sets~\cite{Gomes}.

Through this paper, we aim to expand the study of DLSB on the measurements of oscillation probability in future long baseline (LBL) neutrino experiments, including Deep Underground Neutrino Experiment (DUNE). In order to account for the particularities which separate the SBL and LBL oscillation circumstances, we must make key deviations from the Lagrangian calculation performed by Gomes and Neves. These steps are outlined and justified in Section~\ref{theory}.

Then, in Section~\ref{data analysis}, the resulting transition probabilities from muon neutrinos (antineutrinos) to electron neutrinos (antineutrinos) is studied. In particular, we study the difference between the oscillation probability for cases with and without DLSB. Subsequently, we calculate the minimum number of years of operation for DUNE to measure the effect of DLSB through event reconstruction. Using these results, we evaluate if the region of the DLSB parameter space where the reconciliation of Super-Kamiokande and LSND data occurs, as suggested by Gomes and Neves, can be measured at DUNE to any confidence. Finally, Section~\ref{conclusion} presents the summary and conclusions, including the discussion of the impact of DLSB on the measurement of CP-violations at DUNE, using predicted neutrino energy spectra.

\section{Theory} \label{theory}

\subsection{Adjustments to DLSB Model for LBL Scenario}

The dynamical Lorentz symmetry breaking in a four-Majorana fermion model is presented by Gomes and Neves~\cite{Gomes}. This model involves two fermion fields that correspond to a light left-handed Majorana particle, $\psi_1$, and a heavy right-handed Majorana particle, $\psi_2$. Their masses, $m_1$ and $m_2$, are determined through the type II seesaw mechanism, so we assume $m_2 \gg m_1$ throughout this investigation.

There are important considerations that must be taken as to where our analysis will differ from the referent study. The initial DLSB paper was specifically interested in neutrino oscillation at the short baseline, typically restricted to a neutrino energy per propagation difference of $1$~GeV per $1$~km. This narrow space allows for convenient approximations which are not applicable in the LBL case. Our two fundamental changes are therefore to extend the Gomes and Neves model from two to three flavor neutrino oscillation, and to introduce a non-negligible matter-potential term in response to the higher neutrino energies and greater propagation distances respectively.

The matter effect occurs due to electron forward scattering in a medium of varying electron density (ie: the Earth's crust). As a result, the neutrino mixing angles and the mass eigenstates are altered in in the presence of matter. Consequently, the final transition probability must be re-evaluated. In SBL experiments, the matter effect is negligible as neutrinos rarely interacts with other particles and propagation distances are relatively short. However, for LBL, because the neutrinos travels a large distance through the medium, the total amount of interaction becomes significant. Therefore, the modification is necessary to study the neutrino oscillations at DUNE, where neutrinos constantly interact with the Earth while traveling 1300km until they reach the far detector. 

Furthermore, considering three flavors of neutrino states gives rise to a CP-violation term, $\delta_{cp}$. Incorporating CP-violation into the fermion model allows for studying the difference in oscillation probability of neutrinos and antineutrinos. Consequently, the result of this investigation suggests another factor: namely that DLSB may need to be considered in future oscillation experiments, like DUNE, in measurements of matter-antimatter asymmetry. DLSB oppositely affects the oscillations of neutrinos and antineutrinos, in effect exacerbating the asymmetry. Therefore, we aim to determine how the presence of DLSB affects the measurement of CP-violation at DUNE. In addition, we investigate whether the suggested value of the DLSB parameter from Reference~\cite{Gomes} is measurable at DUNE.

\subsection{Calculations}

The Lagrangian presented by Gomes and Neves is as follows:
\begin{eqnarray}
\mathcal{L}_{model} = && \overline{\psi}_1(i\cancel{\partial}-m_1)\psi_1 - \frac{G_1}{2}(\overline{\psi}_1\gamma_\mu\gamma_5{\psi}_1)^2\nonumber \\
&&+ \overline{\psi}_2(i\cancel{\partial}-m_2)\psi_2 - \frac{G_2}{2}(\overline{\psi}_2\gamma_\mu\gamma_5{\psi}_2)^2\nonumber \\
&& + G_3(\overline{\psi}_1\gamma_\mu\gamma_5{\psi}_1)(\overline{\psi}_2\gamma_\mu\gamma_5{\psi}_2).
\label{orgmodel}
\end{eqnarray}

A small mixing angle is assumed, so $ \cos\theta \simeq 1 $ and $ \sin\theta \simeq 0 $, which allows LHN and RHN states decouple. Consequently, the $\psi_1$ state corresponds to an LHN and, the $\psi_2$ state corresponds to an RHN.

Introducing the auxiliary fields $A^{\mu}$ and $B^{\mu}$, the Lagrangian can be written as

\begin{eqnarray}
\mathcal{L}_{model} = && \overline{\psi}_1(i\cancel{\partial}-g\cancel{A}\gamma_5-m_1)\psi_1 - \frac{G_1}{2}(\overline{\psi}_1\gamma_\mu\gamma_5{\psi}_1)^2\nonumber \\
&&+ \overline{\psi}_2(i\cancel{\partial}-g'\cancel{B}\gamma_-m_2)\psi_2 - \frac{G_2}{2}(\overline{\psi}_2\gamma_\mu\gamma_5{\psi}_2)^2\nonumber \\
&& + \frac{1}{2}g^2_1A_{\mu}A^{\mu} + \frac{1}{2}g^2_2B_{\mu}B^{\mu} + g^2_3A_{\mu}B^{\mu}
\label{orggauge}
\end{eqnarray}
where $g, g'$ and $g_i (i=1,2,3)$ are the constants derived from $G_1,G_2$ and $G_3$.

If the quartic self-couplings and the interaction between the light and heavy particles are considered up to second order, the VEV of the model becomes two 4-vectors, $a^{\mu}$ and $b^{\mu}$. This result indicates the Lorentz symmetry breaking.
In this investigation, we set $a^{\mu} \neq 0$ and $b^{\mu} = 0$, for simplicity. This choice corresponds to assuming that only one auxiliary field, $A^{\mu}$, which is associated with LHN, introduces DLSB. Since $m_2$ is set to be very large, the impact of small VEV will be insignificant compared to the case of LHN, so this is a reasonable assumption. In addition, the DLSB parameter is considered to be time-like,  $a^{\mu} = (a^0,0)$. This setting is equivalent to choosing the preferred direction in spacetime \cite{Carroll}. 

The dispersion relation for the Majorana fermions of this model can be obtained as

\begin{subequations}
\begin{equation}
(p^2-m_1^2-a^2)^2 + 4[p^2a^2-(p\cdot a)^2] = 0,\end{equation}
\begin{equation}
p^2 = m_2^2 \end{equation}
\end{subequations} 

where $a$ is the DLSB parameter proportional to $a^{\mu}$. As a result, the neutrino frequency solution is given as
\begin{eqnarray}
\omega^{(\pm)}_1(\boldsymbol{p}) =&& \sqrt{(|\boldsymbol{p}|\pm a_o)^2 +m_1^2} \nonumber \\
&& \simeq ||\boldsymbol{p}| \pm a_0| + \frac{m_1^2}{2||\boldsymbol{p}| \pm a_o|}
\label{orgfrq}.
\end{eqnarray}
Using Equation \ref{orgfrq}, we can calculate the transition probability of the neutrino model with DLSB. Complete calculations follow identically from Reference~\cite{Gomes}. 

This value can be substituted into the probability equation
\begin{eqnarray}
P(x) = \sin^2(2\theta_{ij}) \sin^2[(E^{(\pm)}_i-E^{(\pm)}_j)\frac{x}{2}],
\label{probability}
\end{eqnarray}
where $E^{(\pm)}_{i}$ represent the energy of $i$th LHN (antineutrino), $\theta_{ij}$ is the mixing angle, $x$ is the travel distance of the neutrinos and $m_{i,j}$ represent mass eigenvalues of LHNs.

By Equation \ref{orgfrq}, $E^{(\pm)}_{i,j}$ in the presence of DLSB becomes
\begin{eqnarray}
E^{(\pm)}_i(\boldsymbol{p}) =&& \sqrt{(|\boldsymbol{p}|\pm a_o)^2 +m_i^2} \nonumber \\
&& \simeq ||\boldsymbol{p}| \pm a_0| + \frac{m_i^2}{2||\boldsymbol{p}| \pm a_o|}
\label{E_i}
\end{eqnarray}
and we can write the value of $E^{(\pm)}_1 - E^{(\pm)}_2$ as
\begin{eqnarray}
E^{(\pm)}_i - E^{(\pm)}_j\simeq \pm (a_i-a_j)+\frac{\Delta m^2_{ij}}{2|E|}
\label{delE},
\end{eqnarray}
assuming $E\gg(a_i,a_j)$. From comparing Equation \ref{delE} with the transition probability equation without considering DLSB, we notice that $\frac{\Delta m^2_{ij}}{2|E|}$ is replaced by $(a_i-a_j)+\frac{\Delta m^2_{ij}}{2|E|}$. This replacement indicates the shift of oscillation probability in the x-axis. 


In addition to the above formalism, the matter effect should be considered for the long-baseline accelerator or solar neutrinos. As these neutrinos propagate through matter, the electron neutrinos experience electron forward scattering causing a phase difference and a change in effective mass \cite{MSW}. In order to consider the matter effect for DLSB, we add the matter potential term to the Lagrangian in Equation \ref{orggauge}. Then, we obtain the modified two Majorana fermions model as follows:
\begin{eqnarray}
\mathcal{L}_{model} = && \overline{\psi}_1(i\cancel{\partial}-g\cancel{A}\gamma_5-m_1-V)\psi_1 - \frac{G_1}{2}(\overline{\psi}_1\gamma_\mu\gamma_5{\psi}_1)^2\nonumber \\
&&+ \overline{\psi}_2(i\cancel{\partial}-g'\cancel{B}\gamma_5-m_2)\psi_2 - \frac{G_2}{2}(\overline{\psi}_2\gamma_\mu\gamma_5{\psi}_2)^2\nonumber \\
&& + \frac{1}{2}g^2_1A_{\mu}A^{\mu} + \frac{1}{2}g^2_2B_{\mu}B^{\mu} + g^2_3A_{\mu}B^{\mu}
\end{eqnarray}
where $V_e=\sqrt{2}G_Fn_e$ is the matter potential. Only left-handed neutrinos, $\psi_1$, interact with electrons in the matter, so only the first term is changed.

The corresponding effective potential is
\begin{eqnarray}
V_{eff}(A,B) = && -\frac{1}{2}g^2_1A^2-\frac{1}{2}g^2_2B^2-g_3^2A\cdot B \nonumber \\
&&+ i\int \frac{d^4p}{(2\pi)^4} tr[\ln(\cancel{p}-m_1-g\cancel{A}\gamma_5-V)] \nonumber \\
&&+ i\int \frac{d^4p}{(2\pi)^4} tr[\ln(\cancel{p}-m_2-g'\cancel{B}\gamma_5)]
\end{eqnarray}

Through the evaluation of the VEV, we find that the new DLSB parameter can be written as
\begin{eqnarray}
a^2 = && 3\pi^2\left[\frac{g^2_{1R}}{g^2}+\frac{m^2_1}{\pi^2}\ln(\frac{m_1}{\Lambda})\right] \nonumber \\
&& \simeq 3\pi^2\frac{g^2_{1R}}{g^2}.
\label{dlsb}
\end{eqnarray}
\\
Since LHNs are assumed to be very small, we use $g^2_{1R} \gg m_1^2$.
Then we define the new corresponding renormalization coupling constant as
\begin{eqnarray}
\frac{g^2_{1R}}{g^2}=\frac{g^2_1}{g^2}-\frac{(m_1+V)^2}{\pi^2\epsilon}.
\end{eqnarray} 
\\
Then we use the formula for oscillation and survival probability from muon neutrino to electron neutrino for three flavors in the presence of matter formulated by Hiroshi Nunokawa and Stephen Parke and José W.F. Valle \cite{Probability}. 
\begin{widetext}
\begin{eqnarray}
\label{generalOSC}
P_{\nu_{\mu} \to \nu_e}(x) \simeq && \sin^2(\theta_{23}) \sin^2 (2\theta_{13}) \frac{\sin^2(\Delta_{31}-V'x)}{(\Delta_{31}-V'x)^2}\Delta_{31}^2 \nonumber \\
&&+\sin(2\theta_{23})\sin(2\theta_{13})\sin(2\theta_{12})\frac{\sin(\Delta_{31}-V'x)}{\Delta_{31}-V'x}\Delta_{31}\frac{\sin(V'x)}{V'x}\Delta_{21} \cos(\Delta_{31}-\delta_{cp})\nonumber \\
&&+ \cos^2(\theta_{23})\sin^2(2\theta_{12})\frac{\sin^2(V'x)}{(V'x)^2}\Delta_{21}^2
\end{eqnarray}

\begin{eqnarray}
\label{survival}
P_{\nu_{e} \to \nu_e}(x) \simeq && 1-4\cos^2(\theta_{12}) \cos^2 (\theta_{13}) \sin^2(\theta_{12}) \cos^2 (\theta_{13}) \frac{\sin^2(\Delta a_{21}-V'x)}{(\Delta a_{21}-V'x)^2}\Delta_{21}^2 \nonumber \\
&&-4 \sin^2(\theta_{12}) \cos^2 (\theta_{13}) \sin^2(\theta_{13}) \sin^2(\Delta a_{32}-\Delta_{32}x) \nonumber \\
&&-4 \sin^2(\theta_{13}) \cos^2(\theta_{12}) \cos^2 (\theta_{13}) \frac{\sin^2(\Delta_{31}+\Delta a_{31}-V'x)}{(\Delta_{31}+\Delta a_{31}-V'x)^2}\Delta_{31}^2
\end{eqnarray}
for $\Delta_{ij} = \frac{\Delta m^2_{ij}x}{4E}$ and $V'= G_FN_e/\sqrt{2}=\frac{V}{2}$.
\end{widetext}
 For the case of DLSB, we replace $\Delta_{ij}$ with $\Delta a_{ij} +\frac{\Delta m^2_{ij}x}{4E}$ as previously discussed. 

We can evaluate $\Delta a_{ij} =(a_i-a_j)$ through expression stated in Equation \ref{dlsb}.
\begin{eqnarray}
\Delta a_{ij} = \frac{\sqrt{3}\pi}{g} && \Delta g^2_{1R} \nonumber \\ 
 = \sqrt{3}\pi[&&\sqrt{\frac{(g_1)_i^2}{g^2}-\frac{(m_i+V)^2}{\pi^2\epsilon}} \nonumber \\ 
&& -\sqrt{\frac{(g_1)_j^2}{g^2}-\frac{(m_j+V)^2}{\pi^2\epsilon}}]
\label{dela}
\end{eqnarray}

Assuming that $g_1 \gg V, m_{i,j}$, which is valid for the $\frac{\Delta(g_1)_{ij}}{g} \sim 10^{-2}-10^{-4}$ GeV, corresponding to strong coupling regime (SCR), Equation \ref{dela} becomes 
\begin{eqnarray}
\Delta a_{ij} = \sqrt{3} \pi \frac{\Delta(g_1)_{ij}}{g}.
\end{eqnarray}
Therefore, we can substitute
\begin{eqnarray}
\Delta_{ij}=\sqrt{3} \pi \frac{\Delta(g_1)_{ij}}{g}+\frac{\Delta m^2_{ij}x}{4E} 
\end{eqnarray}
into Equation~\ref{generalOSC} to obtain the transition probability from electron neutrino to muon neutrino in the presence of DLSB. Through similar steps, we obtain the survival probability of muon neutrinos as Equation~\ref{survival}.

\section{Data Analysis} \label{data analysis}

In this section, we discuss the oscillation probability in the presence of DLSB and the required years of operation at DUNE to detect the differences using the predicted spectra obtained from the event reconstruction. In addition, we study the effect of the DLSB on the measurement of the CP violation at DUNE. 

\subsection{\label{DA:simulation} Reproducing DUNE Simulation}

DUNE is a future neutrino experiment in which the neutrino beam produced inside a particle accelerator, PIP-II, travels 1300 Km between the near detector located at the Fermilab site in Illinois and the far detector located at the Sanford Underground Research Facility in South Dakota. The neutrino beam can be produced in either neutrino or antineutrino mode. The path traveled by the neutrino beam consists of the earth, which interacts with the neutrino beam to change the oscillation rate. The oscillation probability from muon neutrino to electron neutrino can be determined using the neutrino interaction rate data from the near detector and the far detector. By comparing the neutrino and antineutrino oscillation rates, matter-antimatter asymmetry can be studied. 

DUNE is planned to start with a 10 kt far detector mass with a 1.07MW proton beam in the first year, and after a staged construction it is expected to comprise a 20 kt far detector module with a 2.14MW beam in the 7th year of running. The more detailed schedule for the deployment plan can be read in the DUNE conceptual design report~\cite{DUNE}. DUNE is currently scheduled to run for 20 years, resulting in a total exposure of 1326 kt-MW-yrs. The expected total exposure reached by each DUNE operation year is presented in Table \ref{table:exposure}.

\begin{table}
\begin{ruledtabular}
\begin{center}
\begin{tabular}{p{4cm}p{4cm}}
Years & Exposure (kt-MW-yr)\\
\hline
1 & 10.7 \\
5 & 150 \\
8 & 364 \\
13 & 792 \\
20 & 1391 \\

\end{tabular}
\end{center}
\end{ruledtabular}
\caption{\label{table:exposure}%
Total exposure in kt-MW-yr at DUNE after each year of operation according to the deployment plan in DUNE Design Report~\cite{DUNE}. }
\end{table}

The parameter values used for the analysis are presented in Table~\ref{table:parameters}. The parameters related to neutrino oscillations, such as $\Delta m_{31}^2$, $\Delta m_{12}^2$, $\theta_{23}$ and $\theta_{31}$ are determined in M. C. Gonzalez-Garcia, Michele Maltoni and Thomas Schwetz through global fit analysis considering solar, atmospheric, reactor and accelerator data \cite{GlobalFit}. Kevin J. Kelly and Stephen J. Parke show that the density profile of DUNE can be assumed to be constant along the beam line \cite{Density}. This result is supported by the evaluation of the impact on the oscillation probability by the difference in the matter density, using the density models presented by Byron Roe~\cite{PhysRevD.95.113004}. Therefore, for the calculation of the matter potential, we assume that the matter density of $\rho_{Avg} = 2.845 g/cm^3$ and the electron fraction of $Y_e = \frac{1}{2}$. 

\begin{table}
\begin{ruledtabular}
\begin{tabular}{p{2.7cm}p{2.8cm}p{2.5cm}}
Parameter & Value & Uncertainty \\
\hline
$\Delta m_{31}^2$ (NH) & $2.457 \times 10^{-3} eV^2$  & 2.0\% \\
$\Delta m_{31}^2$ (IH) & $-2.449 \times 10^{-3} eV^2$  & 1.9\% \\
$\Delta m_{21}^2$ & $7.5 \times 10^{-5} eV^2$ & 2.4\% \\
$\theta_{12}$ & 0.5843 & 2.3\% \\
$\theta_{13}$ (NH) & 0.738 & 5.9\% \\
$\theta_{13}$ (IH) & 0.864 & 4.9\% \\
\end{tabular}
\end{ruledtabular}
\caption{\label{table:parameters}%
The values and uncertainty of the oscillation parameters obtained from global fit analysis~\cite{GlobalFit}. 
}
\end{table}

\begin{figure} [b]
\includegraphics[width=0.48\textwidth]{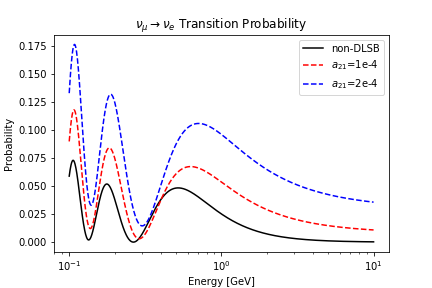}
\quad
\includegraphics[width=0.48\textwidth]{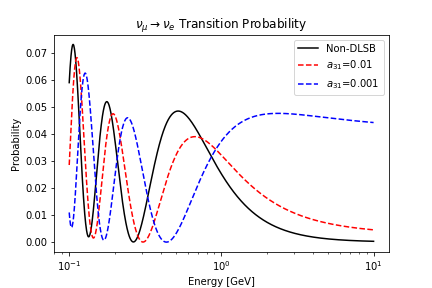}
\caption{\label{fig:sample} The transition probability $P_{\nu_e\to \nu_{\mu}}$ as function of energy E of neutrino in GeV. The travel distance is equal to 1300 Km. The upper panel describes the change in transition probability when changing the DLSB parameter $a_{21}$, and the lower panel describes the change in transition probability $a_{31}$. The other parameter values are used as indicated in Table \ref{table:parameters}.}
\end{figure}

Figure~\ref{fig:sample} describes the transition probability $P_{\nu_{\mu}\to \nu_e}$ in the presence of the matter effect and DLSB as a function of the energy of neutrinos. We set the travel distance $x$ as 1300Km. The two DLSB parameters $a_{21}$ and $a_{31}$ appear in Equation \ref{generalOSC} as a result of extending the theory to three flavor states from only two flavor states. These parameters describe the difference between the VEV of the auxiliary fields of each neutrino state. We study these parameters separately in Figure \ref{fig:sample} to understand each of their effects on the transition probability. 

As illustrated in Figure \ref{fig:sample}, the change in oscillation probability is in the same direction as the differential in $a_{21}$. In contrast, the effect of $a_{31}$ shifts the oscillation probability in the x-axis. This outcome resembles the result of Reference~\cite{Gomes}. Noticeably, not necessarily the higher value of$a_{31}$ results in a larger change in oscillation probability. Certain values of $a_{31}$, such as $a_{31}=0.001$, shift the oscillation probability, so it becomes completely out of phase and causes neutrino flavour transition even at high-energy regions. Conversely, the oscillation probability, for $a_{31}=0.01$, is nearly in phase with the non-DLSB oscillation and generates smaller changes to the oscillation probability. For both $a_{21}$ and $a_{31}$, the matter effect produces the attenuation of the oscillation probability for the larger neutrino energy, in contrast to the oscillations in the vacuum, which can be seen obtained by Gomes and Neves \cite{Gomes}. The studied DLSB parameter values corresponds to SCR, as $\sqrt{3} \pi \frac{\Delta(g_1)_{ij}}{g} \gg \frac{\Delta m^2_{12}x}{4E}$, where $\Delta m^2_{12} \sim 10^{-5}$. 

We examine the sensitivity of DUNE for detecting the effect of the DLSB parameter $a_{21}$ and $a_{31}$. We use the flux at the far detector provided in Reference~\cite{Abi2020} to reconstruct the event rate taking DLSB into account. We obtain the initial flux of muon and electron neutrinos and antineutrinos by calculating the population that survived and went through a transition from each of the neutrinos species for each 0.25 eV of neutrino energy. We use the oscillation probability without DLSB, by setting $a_{21},a_{31}=0$ in Equation \ref{generalOSC}, to do the calculation for the reverse process. Then by changing the values of $a_{21}$ and $a_{31}$ for Equation \ref{generalOSC} and Equation \ref{survival}, we calculate the numbers of oscillated and remained neutrinos after arriving at the far detector. As a result, we obtain the prediction of the spectra at DUNE when DLSB is considered, which are compared with the standard predicted spectra without DLSB. These two models are obtained from the same initial flux of neutrinos, but two different oscillation processes are applied. We then study for which values of $a_{21}$ and $a_{31}$ the two models lead to significant differences that can be measured at DUNE. To compare the effect of DLSB and the CP violation, we also predict the event rate for the case where DLSB parameters are set to 0 and $\delta_{CP}=\pi/2$. 

Figure \ref{Fig:a21} and \ref{Fig:a31} presents the number of events for each neutrino energy after operating DUNE for five years for the two models. For the DLSB model, we study the three values of DLSB parameters $a_{21}$ and $a_{31}$, $0.001$ (Blue), $0.0005$ (Red) and $0.0001$ (Yellow) in the SCR. The effect of $a_{21}$ and $a_{31}$ are studied individually and illustrated in separate plots. Each of the histograms is labeled with the corresponding DLSB parameter values. These hypotheses assuming DLSB are tested against the null hypothesis (Black), which does not take DLSB into account. We label the null hypothesis as "null" and present its statistical uncertainty as given in  Reference~\cite{Abi2020}. If the deviation of the number of events for the DLSB parameter is large enough, the effect of DLSB can be measured at DUNE. We assume $\delta_{cp}=0$ for all the cases except where explicitly stated, as in the gray histogram in Figure~\ref{Fig:a21} and \ref{Fig:a31} where we illustrate the maximum CP violation allowed in a non-DLSB hypothesis, $\delta_{cp}=\pi/2$.

The effect of DLSB is more evident in the electron neutrino (antineutrino) events than the muon neutrino (antineutrino). The much larger initial flux of muon neutrinos than the electron neutrinos results in a larger statistical uncertainty, making the effect of the transition less visible. Notably, the number of events for $a_{31} = 0.001$ deviates exceptionally from the non-DLSB plot compared to the case of two other $a_{31}$ values for electron neutrinos and electron antineutrinos. This is the consequence of the transition probability for $a_{31} = 0.001$ being out of phase with the non-DLSB as described in Figure \ref{fig:sample}. As a result, if $a_{31}=0.001$, the effect of DLSB can be measured at DUNE after the short period of operation.

\begin{widetext}

\begin{figure}
\includegraphics[width=0.4\textwidth]{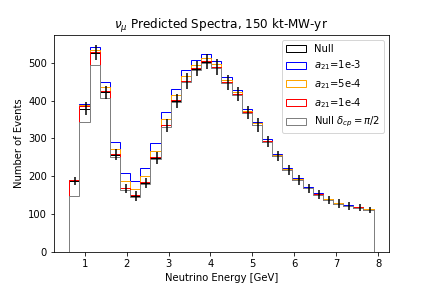}
\includegraphics[width=0.4\textwidth]{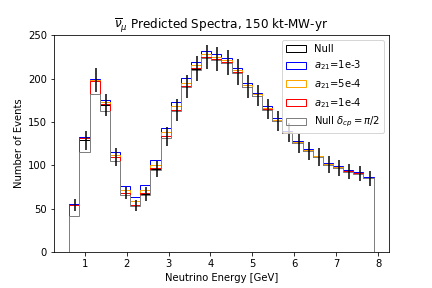}
\end{figure}
\begin{figure}
\includegraphics[width=0.4\textwidth]{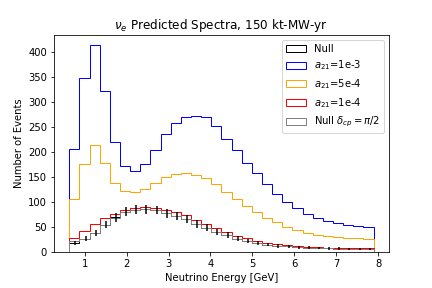}
\includegraphics[width=0.4\textwidth]{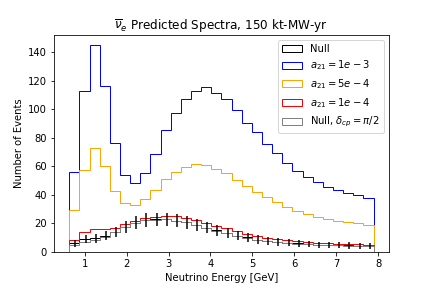}
\caption{\label{Fig:a21} The predicted spectra of $\nu_{\mu}, \overline{\nu}_{\mu},{\nu}_e$ and $\overline{\nu}_e$ after 150 kt-MW-yr of exposure, which can be obtained after operating DUNE for 5 years. We use the flux at the far detector provided in Reference \cite{Abi2020} as the null hypothesis (Black). For DLSB hypothesis, three different values of DLSB parameter $a_{21}$, $0.001$ (Blue), $0.0005$ (Red) and $0.0001$~(Yellow) is illustrated. We set $\delta_{cp}=0$ for all the events, except for the non-DLSB case with $\delta_{cp}=\pi/2$ (grey). The statistical error for the null hypothesis is included, as given in Reference \cite{Abi2020}.}
\end{figure}
\begin{figure}
\includegraphics[width=0.4\textwidth]{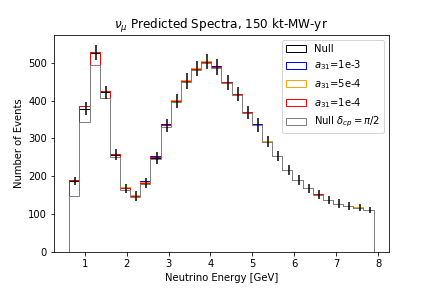}
\includegraphics[width=0.4\textwidth]{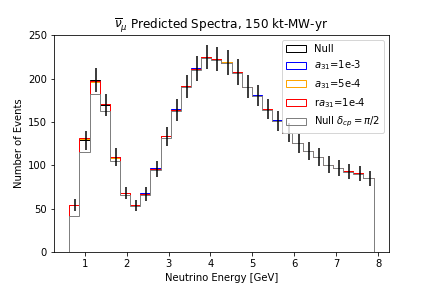}
\includegraphics[width=0.4\textwidth]{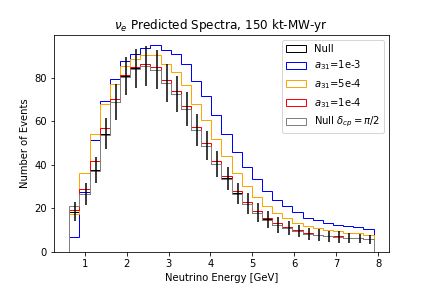}
\includegraphics[width=0.4\textwidth]{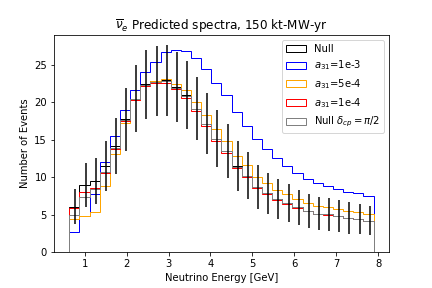}
\caption{\label{Fig:a31} The predicted spectra of $\nu_{\mu}, \overline{\nu}_{\mu},{\nu}_e$ and $\overline{\nu}_e$ after 150 kt-MW-yr of exposure which can be obtained after
operating DUNE for 5 years. We use the flux at the far detector provided in Reference \cite{Abi2020} as the null hypothesis (Black). For DLSB hypothesis, three different values of DLSB parameter $a_{31}$, $0.001$ (Blue), $0.0005$ (Red) and $0.0001$~(Yellow) is illustrated. We set $\delta_{cp}=0$ for all the events, except for the non-DLSB case with $\delta_{cp}=\pi/2$ (grey). The statistical error for the null hypothesis is included, as given in Reference \cite{Abi2020}.}
\end{figure}
\end{widetext}

\subsection{\label{DA:chi2} Comparing Spectra}
\begin{figure}
\includegraphics[width=0.48\textwidth]{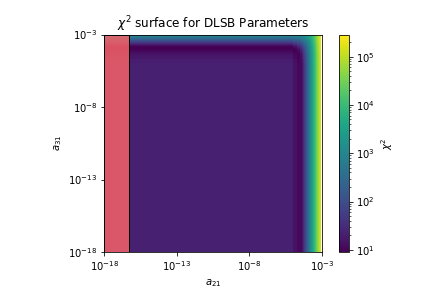}
\includegraphics[width=0.48\textwidth]{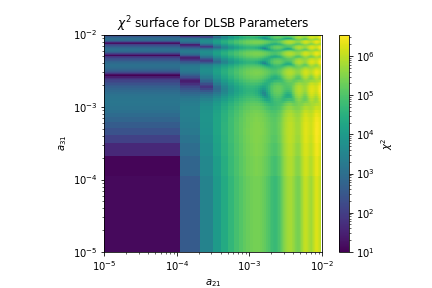}
\caption{\label{fig:chi2} $\chi^2$ surface for the two DLSB parameters, $a_{21}$ and $a_{31}$, obtained by comparing the event rate prediction with corresponding DLSB values to non-DLSB spectra for each 0.25 GeV neutrino energy after operating DUNE for 20 years. The color axis represents the $\chi^2$, where yellow corresponds to high $\chi^2$ values. The highlighted region (red) is for $a_{21}$ value that reconciles the super-Kamiokande and LSND data, suggested by Gomes and Neves \cite{Gomes}.}
\end{figure}

Then we perform the $\chi^2$ analysis on the predicted spectra of DLSB oscillation and non-DLSB oscillation from Reference ~\cite{Abi2020}. We compare the predicted oscillation taking DLSB into account with a null hypothesis as detailed in the previous section. We choose the degrees of freedom to be 1, as two variables, $a_{21}$ and $a_{31}$, are changed between the two models. We present the $\chi^2$~surface of $a_{21}$ and $a_{31}$ between the rage $10^{-19} - 10^{-3}$ (upper) and $10^{-5} - 10^{-3}$ (lower) for the case of electron neutrino oscillation after operating DUNE for 20 years in Figure~\ref{fig:chi2}. The colored axis represents the $\chi^2$ values. Yellow represents the areas of lower agreement between the two models where DLSB would be easy to observe. 

Up to $10^{-3}$, spectra with higher values of DLSB parameters have high $\chi^2$ values, as expected. It is noticeable that there is fluctuation of $\chi^2$ values in the region between 0.001 and 0.01 for both $a_{21}$ and $a_{31}$ values. As discussed in the previous section, this results from certain DLSB parameter values result in generating oscillations in phase with non-DLSB oscillation. Within the region of DLSB parameters displayed in Figure~\ref{fig:chi2}, $a_{21} = 0.00971$ and $a_{31}=0.00171$ result the maximum $\chi^2$. Also, among the two DLSB parameters, $a_{21}$ has a larger impact on the visibility of the DLSB effect at DUNE.

We indicate in red the region $a_{21} \sim \pm 6 \sqrt{3} \pi \times 10^{-19}$, in which the super-Kamiokande and LSND data can be reconciled, suggested by Gomes and Neves in the upper panel of Figure~\ref{fig:chi2}. Most of the highlighted range is inside the region of low $\chi^2$, meaning that it is very hard for DUNE to observe this value of DLSB, except for the case where $a_{31}$ is in much higher order than $a_{21}$. 

We also present the minimum years of operation required to measure DLSB with 99\% confidence for each $a_{21}$ and $a_{31}$ value in Table \ref{table:required years}. Starting from the predicted event rate at 150 kt-MW-yr, we increased the exposure up to 1391 kt-MW-yr and accepted the minimum amount of exposure that resulted in a p-value less than 0.01, which means that the predicted plot is improbable to be obtained in the case without DLSB. Then we converted the amount of exposure to years using Table~\ref{table:exposure}. Since the DUNE is planned to operate for 20 years, if the value in this table is lower than 20, the DLSB parameter for the neutrino sample can be observed. In Table~\ref{table:required years}, only one of the DLSB parameters is changed at a time. Therefore the values correspond to the x and y axis of Figure~\ref{fig:chi2}. 

According to Table \ref{table:required years}, the DLSB parameters as small as $10^{-4}$ can be detected at DUNE, except for $a_{31}$ for electron antineutrino, which requires more than 20 years to obtain enough sample. For the case of muon neutrinos and antineutrinos, $a_{21}$ as small as $0.0005$ is detectable, but the effect $a_{31}$ is very hard to be measured even for higher values of 0.001. This result is consistent with the results of Figure \ref{fig:sample}, where $a_{21}$ resulted in larger deviation from non-DLSB oscillations than $a_{31}$. DLSB can generally be more easily detected from the electron neutrino and antineutrino channels than muon channels. The testable region, $a_{21},a_{31} \sim 10^{-3}-10^{-4}$ corresponds to the SCR.

\begin{table}
\begin{ruledtabular}
\begin{tabular} {p{2cm}|p{2cm}p{2cm}p{2cm}}
$a_{31}$ & 0.001 & 0.0005 & 0.0001\\
\hline
$\nu_e$ & 5 & 5 & 13 \\
$\overline{\nu}_e$ & 5  & 5 & 20 $\ll$ \\
$\nu_{\mu}$ & 20 $\ll$ & 20 $\ll$  & 20 $\ll$  \\
$\overline{\nu}_{\mu}$ & 20 $\ll$ & 20 $\ll$ & $20 \ll$ \\
\hline
$a_{21}$ &0.001 &0.0005 &0.0001\\
\hline
$\nu_e$ & 5 & 5 & 5 \\
$\overline{\nu}_e$ & 5  & 5 & 5 \\
$\nu_{\mu}$ & 5 & 5 & 20 $\ll$ \\
$\overline{\nu}_{\mu}$ & 5 & 10 & 20 $\ll$ \\
\end{tabular}
\end{ruledtabular}
\caption{\label{table:required years}%
Required years of operation for DUNE to be able to measure the effect of DLSB from $\nu_e$, $\overline{\nu}_e$, $\nu_{\mu}$ and $\overline{\nu}_{\mu}$ signals for $a_{31} \sim 10^{-3}-10^{-4}$ (Upper) and $a_{21} \sim 10^{-3}-10^{-4}$ (Lower). For each table, the other parameter is assumed to be 0.
}
\end{table}

\section{Conclusion} \label{conclusion}

 Our aims were to determine the orders of magnitude of DLSB that can be measured at DUNE and to find out the impact of DLSB on a LBL measurement of CP-violation -- one of DUNE's major stated goals. We extend the neutrino model with DLSB introduced by Gomes and Neves~\cite{Gomes} in order to incorporates the relevant effects consistent with future LBL experiments. Additionally, we consider all three flavors of neutrinos to calculate the oscillation probability instead of two to reflect the actual interaction in the experiment. It gives rise to two separate DLSB parameters, $a_{21}$ and $a_{31}$, and the CP violation term involved in the calculation of $P_{\nu_e\to \nu_{\mu}}$. The DLSB parameters correspond to the difference between the VEV of the auxiliary field of $i$th neutrino state for $i=1,2,3$ and each of them affects neutrino oscillations distinctively: $a_{21}$ raises or lowers the transition probability at all energy levels, and $a_{31}$ causes shift in energy-over-time axis in Figure~\ref{fig:sample}. 
 
 With the modified oscillation probability, we create predicted spectra for $\nu_e$, $\overline{\nu}_e$, $\nu_{\mu}$, and $\overline{\nu}_{\mu}$ using the data of neutrino flux at far detector from the DUNE design report~\cite{DUNE}. If DUNE collects data for 20 years as currently planned, $a_{21}$ in range $10^{-3} - 10^{-4}$ causes significant difference on the oscillations of $\nu_e$, $\overline{\nu}_e$, $\nu_{\mu}$ and $\overline{\nu}_{\mu}$. $a_{31}$ causes relatively smaller impact on the oscillations, so its impact will be observable especially on $\nu_e$ and $\overline{\nu}_e$ signals. These measurable values of the DLSB parameter have much higher orders of magnitude compared to $\Delta m^2$ or the matter potential, $V$, so they belong to SCR. We plot $\chi^{2}$ surface of DLSB parameters when the neutrino oscillation model incorporating DLSB is tested against the null hypothesis, to determine the values of DLSB parameters that may cause significant changes to the measurement at DUNE. As expected DUNE is more likely to detect observe a DLSB spectrum with higher coupling parameters. Nevertheless, for the region $a_{21},a_{31} \sim 10^{-3}-10^{-2}$, not necessarily the higher parameter values made the effect of DLSB more detectable at DUNE, as certain values of DLSB parameters shifts the oscillation probability to be out of phase with the null model. From the $\chi^{2}$ analysis, we conclude that $a_{21} \sim \pm 6 \sqrt{3} \pi \times 10^{-19}$, the region proposed to reconcile Super-Kamiokande and LSND data by Gomes and Neves in Reference~\cite{Gomes} will not be detectable at DUNE.
 
Another aim of this paper is to study if the DLSB would affect the measurement of the CP violation at DUNE, since DLSB affects the oscillation of neutrinos and antineutrinos in opposite directions. In particular, we examine whether the effect of DLSB resembles the effect of the CP violation, so the measurement at DUNE might become larger than the actual amount of CP violation due to DLSB. We plot the predicted spectra at the far detector for the null oscillation with the maximum CP violation in Figure~\ref{Fig:a31}, in addition to the spectra resulted by the presence of DLSB where CP violation is set to 0. The spectra with DLSB parameters at the detectable range are shown to be distinct from the spectra with maximum CP violation. Therefore, we conclude that DLSB creates an additional systematic uncertainty in the measurement of the CP violation. Therefore, if DLSB corresponding to SCR presents, the DUNE would require a longer operation period to be able to measure the CP violation than the standard case without DLSB. Finally, we compute the best fit DLSB parameter value that minimizes the $\chi^{2}$ when compared to the null oscillations with maximum CP violation for electron neutrino. Within the region displayed in Figure~\ref{fig:chi2}, the DLSB oscillation with $a_{21} \sim 4 \times 10^{-6}$ most closely resembles the effect of CP violation. 

\begin{acknowledgments}
I want to thank Davio Cianci for providing insights and mentorship throughout my research. 
\end{acknowledgments}

\bibliography{aapmsamp}

\end{document}